\documentclass{eas}
\usepackage{graphicx}

\TitreGlobal{Setting a New Standard in the Analysis of Binary Stars}
\begin{document}

\title{Asteroseismology of binary stars and a compilation of core overshoot and
  rotational frequency values of OB stars} \runningtitle{Asteroseismology of
  massive binaries}

\author{Conny Aerts}\address{Institute of Astronomy, KU\,Leuven, B-3001 Leuven,
  Belgium; Department of Astrophysics, IMAPP, Radboud University 
Nijmegen, the Netherlands}

\begin{abstract}
After a brief introduction into the asteroseismic modelling of stars, we provide
a compilation of the current seismic estimates of the core overshooting
parameter and of the rotational frequency of single and binary massive
stars. These important stellar parameters have meanwhile become available for
eleven OB-type stars, among which three spectroscopic pulsating binaries and one
magnetic pulsator.  We highlight the potential of ongoing and future analyses of
eclipsing binary pulsators as essential laboraties to test stellar structure and
evolution models of single and binary stars.
\end{abstract}

\maketitle

\section{The Asteroseismic Modelling of Stars}

\begin{figure}
\begin{center}
\rotatebox{0}{\resizebox{8cm}{!}{\includegraphics{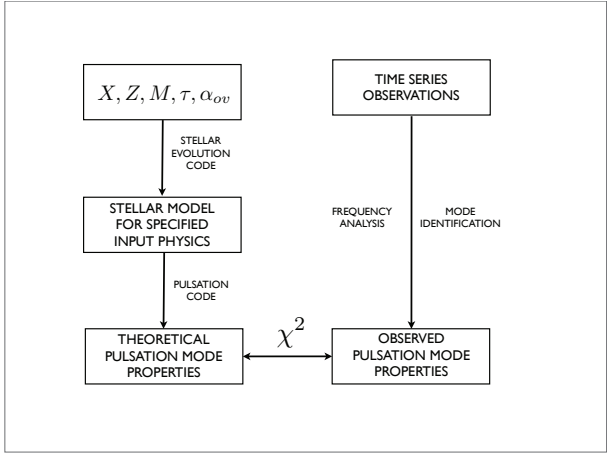}}}
\end{center}
\caption[]{Schematic representation of the procedure of forward seismic
  modelling of a massive pulsator with a convective core; $\tau$ stands for the
  age of the star and $\alpha_{\rm ov}$ for its core overshoot parameter
  expressed in units of the local pressure scale height. Figure courtesy of
  Dr.\ Katrijn Cl\'emer. }
\label{context}
\end{figure}

Asteroseismology is undergoing a revolution since the operation of space
missions dedicated (partly) to this subject, such as MOST, CoRoT, and {\it
  Kepler\/} launched in 2003, 2006, and 2009, respectively.  The past decade has
seen the assembly of uninterrupted white-light space photometry with
$\mu$mag-precision for thousands of stars that turn out to be pulsating.  A
thorough introduction into the field of asteroseismology was recently presented
in the monograph by Aerts et al.\ (2010) and is thus omitted here. The basic
principle of asteroseismic modelling is summarized in one snapshot in the
context diagram shown in Fig.\,\ref{context}.

Following a too simplistic point of view, one could say that current research
in asteroseismology is done with two major aims:
\begin{enumerate}
\item
To deliver high-precision stellar parameters resulting from the scheme in
Fig.\,\ref{context} for exoplanet host stars and for thousands of stars covering
large ranges of mass, age, metallicity, and populations in the Milky Way, as
input for further stellar and galactic studies.  Hereby, it is assumed that the
input physics of the stellar structure models is sufficiently appropriate, just
as this is assumed in the case when classical data are used to deduce stellar
parameters.
\item
To improve the theory of stellar structure and evolution, both for single and
binary stars, by focusing on the shortcomings of the input physics reflected by
too high $\chi^2$-values with respect to the measurement errors.
\end{enumerate}

Even the most basic seismic analysis, based on the frequency of maximum power as
well as on the large frequency separation for solar-like pulsators (e.g.,
Chaplin et al.\ 2011), or on the average period spacing and the periodic
deviations thereof for high-order gravity-mode pulsators (e.g., Degroote et
al.\ 2010), leads to values of the global stellar parameters, such as the mass,
radius, and age, with a relative precision of only a few percent, i.e., far
better than what can be deduced from photometric colour indices, spectral
analysis, interferometric data or present-day astrometry.  As a keynote recent
illustration of this, the combination of frequency and period spacings of the
dipole mixed modes detected in evolved low-mass stars allowed to distinguish
between red giants with only hydrogen-shell burning while 
climbing up the red giant
branch or with core helium burning in addition after the helium flash (Bedding
et al.\ 2011), a probing that cannot be done from  classical data.
\begin{table}
\caption{\label{tabel}Summary of observed ($v\sin\,i$, $T_{\rm eff}$, $\log\,g$,
  $f_{\rm rot}$) and seismically modelled ($M$, $X_c$, $\alpha_{\rm ov}$)
  stellar parameters of 11 OB-type pulsators. The star indicated in bold has a
  magnetic field while the three spectroscopic binaries are indicated in
  italic. The data sources of the observed table entries are listed in Aerts et
  al.\ (2014) and are omitted here for brevity, the references for the seismic
  modelling results are listed in the last column as a number according to the
  footnote.}  \tabcolsep=3pt
\begin{tabular}{ccccccccc}
\hline\hline
\multicolumn{1}{c}{HD number}&
\multicolumn{1}{c}{vsini}&
\multicolumn{1}{c}{$f_{\rm rot}$}&
\multicolumn{1}{c}{$\log\,T_{\rm eff}$}&
\multicolumn{1}{c}{$\log\,g$}&
\multicolumn{1}{c}{$\alpha_{\rm ov}$}&
\multicolumn{1}{c}{Mass}&
\multicolumn{1}{c}{$X_c$}&
\multicolumn{1}{c}{Ref.}\\
\multicolumn{1}{c}{}&
\multicolumn{1}{c}{(km\,s$^{-1}$)}&
\multicolumn{1}{c}{(d$^{-1}$)}&
\multicolumn{1}{c}{(K)}&
\multicolumn{1}{c}{(dex)}&
\multicolumn{1}{c}{($H_{\rm p}$)}&
\multicolumn{1}{c}{($M_\odot$)}&
\multicolumn{1}{c}{($\%$)}&
\multicolumn{1}{c}{}\\
\hline
 16582&  1&  0.075& 4.327& 3.80& $0.20\pm0.05$& 10.2& 0.25&(1)\\
 29248&  6&  0.017& 4.342& 3.85&  $<$0.12    & 9.5 &0.26&(2)\\
 44743& 23&  0.054& 4.380& 3.50& $0.20\pm0.05$& 13.6& 0.12&(3)\\
 46202& 25&  ---  & 4.525& 4.10& $0.10\pm0.05$& 24.0& 0.58&(4)\\
129929&  2&  0.012& 4.389& 3.95& $0.10\pm0.05$&  9.4& 0.35&(5)\\
{\bf 163472}& 63&  0.275& 4.352& 3.95&  $<$0.15    & 8.9& 0.29&(6)\\
180642& 25&  0.075& 4.389& 3.45&  $<$0.05    &11.6& 0.23&(7)\\
214993& 36&  0.120& 4.389& 3.65&  $<$0.40    &12.2& 0.28(8)\\
\hline
{\it  50230}&  7&  0.044& 4.255& 3.80& $0.25\pm0.05$& 7.5& 0.28&(9)\\
{\it  74560}& 13&  0.010& 4.210& 4.15&  $<$0.10     &--- &---&(10)\\
{\it 157056}& 31&  0.107& 4.398& 4.10& $0.44\pm0.07$& 8.2& 0.38&(11)\\
\hline
\end{tabular}

{\small (1) Aerts et al.\ (2006); (2) Pamyatnykh et al.\ (2004); (3) Mazumdar et
  al.\ (2006); (4) Briquet et al.\ (2011); (5) Dupret et al.\ (2004); (6)
  Briquet et al.\ (2012); (7) Aerts et al.\ (2011); (8) Desmet et al.\ (2009);
  (9) Degroote et al.\ (2010); (10) Walczak et al.\ (2013); (11) Briquet et
  al.\ (2007).  }
\end{table}

\begin{figure}[ht!]
\begin{center}
\rotatebox{270}{\resizebox{8.cm}{!}{\includegraphics{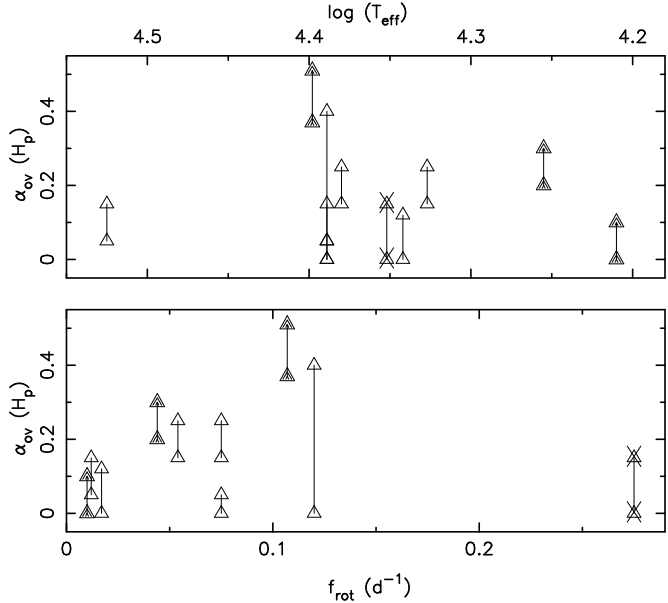}}}
\end{center}
\caption[]{The core overshoot parameter of OB pulsators as a function of the
  observed values for the rotational frequency and effective temperature. Open
  symbols are single pulsators and filled symbols represent spectroscopic
  binaries with a pulsating component. The cross indicates a magnetic pulsator.}
\label{alpha_obs}
\end{figure} 
\begin{figure}[ht!]
\begin{center}
\rotatebox{270}{\resizebox{8.cm}{!}{\includegraphics{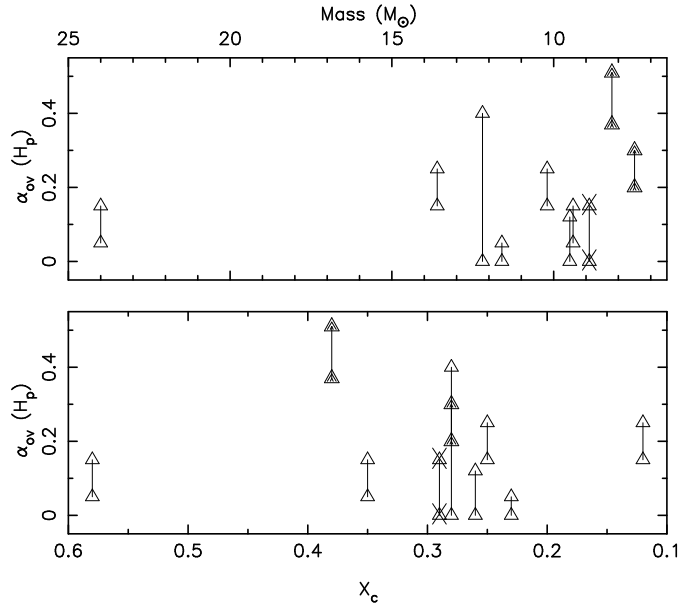}}}
\end{center}
\caption[]{The core overshoot parameter of OB pulsators as a function of the
central hydrogen fraction and the mass as determined from seismic modelling.
The symbols have the same meaning as in Fig.\,\ref{alpha_obs}.}
\label{alpha_theo}
\end{figure}

So far, the main focus of asteroseismology has been put on item 1.\ above, i.e.,
on the derivation of basic stellar parameters, assuming that the input physics
of the theoretical stellar models, as indicated in the left part of
Fig.\,\ref{context}, is correct.  This was of course the first thing to do after
high-precision data came in.  The largest benefit of asteroseismic modelling is,
however, yet to come. It requires the detailed modelling of all the individual
detected and identified oscillation modes in carefully selected stars of various
kinds, in terms of the assumed input physics, with the aim to improve the latter
(item 2.\ above).  Intensive future efforts on the left part of
Fig.\,\ref{context} are necessary to achieve this, based on joint collaborations
between asteroseismologists and experts in the theory of stellar structure.
Progress will be made in the next few years by studying the impact of changes in
various aspects of the input physics on the oscillation properties, just as it
was done in helioseismology to get a better model of the Sun (e.g.,
Christensen-Dalsgaard 2002 for a thorough review). For the moment, we are not
yet at that stage for stars, given the focus on the observational aspects of
asteroseismology during the past few years.

\section{Compilation of Seismic Analysis Results of OB-type Stars}

In an attempt to offer new tools to evaluate theoretical stellar models of
massive stars in terms of interior mixing processes, Aerts et al.\ (2014) made a
compilation of 68 OB-type nearby stars undergoing the CNO cycle in their
convective core and studied their observational properties, including
oscillations, rotation, magnetic field, and nitrogen abundance, from careful
multivariate statistical analysis.  This led to the conclusion that the
effective temperature and the frequency of the dominant acoustic mode are
significant predictors for the nitrogen abundance, while the rotation
diagnostics are not.  This result implies that the oscillation properties should
not be ignored in the evaluation of stellar evolution models.

To trigger further asteroseismic studies of massive stars, including eclipsing
binaries, we assembled the eleven stars in the sample by Aerts et al.\ (2014)
for which seismic modelling according to the scheme in Fig.\,\ref{context} has
been successful and led to a derivation of either a value or an upper limit for
the core overshoot parameter $\alpha_{\rm ov}$, based on the Schwarzschild
criterion of convection and assuming a fully mixed overshoot region. The
spectroscopic and seismic properties of those stars, which are all slow
rotators, are listed in Table\,\ref{tabel}.

Their seismically derived core overshoot parameters are plotted as a function of
two observables, $\log\,T_{\rm eff}$ and $f_{\rm rot}$, in
Fig.\,\ref{alpha_obs}, where the three spectroscopic binaries are indicated with
full symbols and the one magnetic pulsator has an additional cross indication.
By itself, the rotational frequency, which was deduced from Fourier analysis of
the seismic data without any model assumption, is not an obvious predictor for
the amount of core overshooting. A similar conclusion holds for the mass and the
central hydrogen fraction (which is a proxy for the evolutionary state of the
star), as can be seen from Fig.\,\ref{alpha_theo}, although the sample is still
limited, particularly at high masses. Earlier studies suggested that the core
overshoot parameter increases with increasing stellar mass for stars with M$\in
[1.1,1,7]\,$M$_\odot$ (Clausen et al.\ 2010; Torres, these proceedings).
Asteroseismology shows that this conclusion is too simplistic for stars with
masses above 10\,M$_\odot$, in line with the results of Claret (2007) based on
massive eclipsing binaries.

Staritsin (2013) considered nine of the stars in Table\,\ref{tabel} to make
three-dimensional hydrodynamical simulations of the extra mixing at the boundary
of the convective core, based on the physical model of turbulent entrainment
proposed by Meakin \& Arnett (2007). He found that the overshoot parameter
deduced from the simulations decreases as the star moves along its evolutionary
track (cf.\ his Fig.\,3).  Our findings represented in Fig.\,\ref{alpha_theo}
are not in contradiction with this conclusion, as can be deduced from e.g., the
three stars with seismic mass between 8.9 and 9.5\,M$_\odot$, but the sample is
not yet suitable to test this result in detail. Such a test would require
several seismic estimates of $\alpha_{\rm ov}$ for a particular value of the
stellar mass, as a function of $X_c$.

Unfortunately, we have only two seismic values of
$\alpha_{\rm ov}$ for pulsating B stars in close binaries (Table\,\ref{tabel}),
but several new case studies are on the way, such as two SB2 pulsators
discovered from {\it Kepler\/} data (P\'apics et al.\ 2013).
Although the current sample is too small to be
statistically meaningful, this is one of the important and promising ways
towards improving the implementation of the input physics of massive stars.

\section{Eclipsing Binaries: Complications and Opportunties}

Given that both the modelling of eclipsing binaries and the seismic modelling of
stars are two independent methods to deduce interior physics constraints, among
which the overshoot parameter and the age, it is obvious to try and combine
them. This idea is not new but good data to bring it into practice with
predictive power for the improvement of stellar physics had to await
uninterrupted high-precision photometry from space. These data have shown that
asteroseismology of pulsating eclipsing binaries turns out to be far from
trivial, in part because the binary modelling tools were not up to the precision
of the {\it Kepler\/} data (e.g., Degroote, these proceedings). Effects like
Doppler beaming and gravitational lensing occur at measurable amplitudes
reaching several 100$\mu$mag and thus have to be taken into account in the
binary modelling to achieve valid interpretations (e.g., van Kerkwijk et
al.\ 2010; Bloemen et al.\ 2011, 2012). Conversely, the high quality of
the photometric data allows to discover and interpret binarity from careful
pulsational analyses thanks to the detection of the 
R\o mer delay, even without having
spectroscopic data at hand (Shibahashi \& Kurtz 2012, Telting et al.\ 2012).

Both the CoRoT and {\it Kepler\/} missions led to the discovery of numerous
eclipsing binary pulsators, with a variety of flavours in terms of masses and
evolutionary stages of the components (Pr\v{s}a, these proceedings).  An
extensive overview of pulsating binaries with high-precision photometry by {\it
  Kepler\/} will become available in Huber (2014) while several case studies are
discussed elsewhere in the current proceedings. Here, we limit ourselves to
highlight a few remarkable case studies of pulsating eclipsing binaries, without
any effort of being exhaustive and referring to the original papers for details.

The unravelling of pulsational and binary variability, which is a prerequisite
for seismic modelling, necessitates rather complex iterative data-analysis
schemes when the intrinsic and extrinsic variations have about equal amplitude,
such as in the case of the double-lined eclipsing binary KIC\,11285625 analysed
by Debosscher et al.\ (2013). Even if the data analysis can be successfully
accomplished, challenges are faced with the physical interpretation,
particularly in the cases where tidally excited g-mode oscillations occur amidst
(unidentified) free oscillations of (one of) the components in eccentric systems
or when reflection effects are so strong that asymmetrically heated atmosphere
models must be used before seismic interpretations can be attempted. Examples of
eclipsing binaries with a main-sequence pulsator are, e.g., HD\,174884 (Maceroni
et al.\ 2009), CoRoT\,102918586 (Maceroni et al.\ 2013), KIC 10661783
(Southworth et al.\ 2012, Lehmann et al.\ 2013), and KIC\,4544587 (Hambleton et
al.\ 2013). The compact subdwarf pulsator 2M1938+4603 requires a new generation
of atmosphere models before a solid interpretation can be made (\O stensen et
al.\ 2010).  These case studies represent a variety of situations where the
pulsational information was too limited, or was compatible with current-day
evolutionary models of single stars, or requires new theoretical developments in
atmosphere and interior physics.  The latter was the case for the extremely
eccentric F-type binary KOI-42 (Welsh et al.\ 2011), where strong forces due to
dynamical tides trigger nonlinear resonant locking of pulsation modes (Fuller et
al.\ 2012).

Easier cases to analyse and interpret were also found, e.g., KIC\,8410637 which
is a 408-day period eclipsing binary containing a red giant with solar-like
oscillations discovered shortly after the launch of {\it Kepler\/} (Hekker et
al.\ 2010). This object is ideally suited as a test laboratory to evaluate the
scaling relations of solar-like oscillations (e.g., Huber et al.\ 2011) as well
as to assess isochrone fitting in eclipsing binaries and in open clusters
(Frandsen et al.\ 2013). Many other much more complicated red giant binaries
with shorter eccentric orbits have recently been found and are being analysed on
the basis of {\it Kepler\/} photometry combined with long-term follow-up
spectroscopy (Gaulme et al.\ 2013, Beck et al.\ 2014).

\section{Future Studies of Massive Binary Pulsators}

While various studies of pulsating eclipsing binaries based on {\it Kepler\/}
and spectroscopic data are ongoing (e.g., Schmid et al., these proceedings), we
are not aware of any such new case studies with OB-type
primaries. Unfortunately, pulsating OB-type stars in eclipsing binaries are
scarce and the few known ones have insufficient seismic data to tune stellar
physics. This is why a {\it Kepler\/} guest observer proposal focused on the
massive binary V380\,Cyg, the brightest star observed by the {\it Kepler\/}
satellite by means of a dedicated mask.  Tkachenko et al.\ (2012) discovered
low-amplitude photometric and spectroscopic variability in a preliminary
analysis of {\it Kepler\/} photometry and high-resolution spectroscopy. Further
data gathering and analysis delivered radii and masses of a relative precision
near 1\%, but pointed out that the line-profile variations are connected with
spots of Si while the photometric variability is of stochastic nature with
unknown cause (Tkachenko et al.\ 2014). Even without a good explanation of the
intrinsic variability of the primary, it was re-emphasized that single-star
models cannot explain the components of the V380\,Cyg system, not even if we
include a high value for the core overshooting.

Interesting ongoing case studies of double-lined spectroscopic binaries with at
least one pulsating OB component concern Spica (see also K\"onigsberger, these
proceedings) and $\sigma\,$Scorpii, the latter binary's primary being a
large-amplitude radial-mode pulsator for which we recently discovered additional
low-amplitude modes in an extensive data set of high-resolution spectroscopy
(Tkachenko et al., in preparation). So given that the majority of massive stars
are in binaries (de Koter, Sana, these proceedings), we are in need of new
dedicated observing campaigns to improve their modelling via asteroseismology.


\end{document}